\documentclass[a4paper,11pt]{fullverllncs}
\usepackage[left=2.5cm,top=2.5cm,right=2.5cm,centering]{geometry}

\usepackage{graphicx}

\usepackage{amsmath,amssymb}
\usepackage{graphicx}
\usepackage{ascmac}
\usepackage{array}
\usepackage{algpseudocode}
\usepackage{amsfonts}
\usepackage{amssymb}
\usepackage{amsmath}
\usepackage{algorithm, algpseudocode}
\usepackage{multirow}
\usepackage{arydshln}
\usepackage{hhline} 
\usepackage[misc,geometry]{ifsym} 
\usepackage{scalefnt}

\usepackage[dvipdfmx, colorlinks]{hyperref, xcolor}
\definecolor{winered}{rgb}{0.5,0,0}
\definecolor{darkblue}{rgb}{0,0,0.5}
\definecolor{darkgreen}{rgb}{0,0.3,0}
\hypersetup{
linkcolor=winered,
citecolor=darkblue,
urlcolor=darkgreen
}

\newcommand{\Adv}{\mathsf{Adv}}

\newcommand{\pp}{\mathsf{pp}}

\newcommand{\A}{\mathsf{A}}
\newcommand{\B}{\mathsf{B}}

\newcommand{\Z}{\mathbb{Z}}
\newcommand{\G}{\mathbb{G}}

\newcommand{\BG}{\mathsf{BG}}
\newcommand{\Hash}{\mathsf{H}}
\newcommand{\CH}{\mathsf{CH}}
\newcommand{\prfF}{\mathsf{F}_{\mathsf{prf}}}
\newcommand{\prfK}{\mathsf{K}_{\mathsf{F}_\mathsf{prf}}}
\newcommand{\prfout}{\mathsf{out}_{\mathsf{F}_\mathsf{prf}}}
\newcommand{\crs}{\mathsf{crs}}

\newcommand{\pk}{\mathsf{pk}}
\newcommand{\sk}{\mathsf{sk}}

\newcommand{\DID}{\mathsf {DID}}

\newcommand{\gk}{{q, \allowbreak \G_1, \allowbreak \G_2, \allowbreak \G_T, \allowbreak e, \allowbreak G_1, \allowbreak G_2}}

\newcommand{\Fix}{\mathsf{Fix}}
\newcommand{\tlef}{\hspace{-14pt}}
\newcommand{\rHW}{r_\mathsf{HW}}

\newcommand{\Sign}{\mathsf{Sign}}

\newcommand{\Priv}{\mathsf{Priv}}

\newcommand{\DDH}{{\mathrm{DDH}}}
\newcommand{\sfDDHO}{{\mathsf{DDH1}}}

\newcommand{\SXDH}{{\mathrm{SXDH}}}

\newcommand{\qcoDL}{{q\mathrm{\mathchar`-co\mathchar`-DL}}}
\newcommand{\ellcoDL}{{\ell\mathrm{\mathchar`-co\mathchar`-DL}}}
\newcommand{\sfqcoDL}{{q\mathsf{\mathchar`-co\mathchar`-DL}}}
\newcommand{\qcoGSDH}{{q\mathrm{\mathchar`-co\mathchar`-GSDH}}}
\newcommand{\ellcoGSDH}{{\ell\mathrm{\mathchar`-co\mathchar`-GSDH}}}
\newcommand{\sfqcoGSDH}{{q\mathsf{\mathchar`-co\mathchar`-GSDH}}}

\newcommand{\DS}{{\mathsf{DS}}}
\newcommand{\DSSetup}{\mathsf{DS.Setup}}
\newcommand{\DSKeyGen}{\mathsf{DS.KeyGen}}
\newcommand{\DSSign}{\mathsf{DS.Sign}}
\newcommand{\DSVerify}{\mathsf{DS.Verify}}
\newcommand{\pkDS}{\mathsf{pk}_{\DS}}
\newcommand{\skDS}{\mathsf{sk}_{\DS}}
\newcommand{\ppDS}{\mathsf{pp}_{\DS}}
\newcommand{\sfEUFCMA}{\mathsf{EUF\mathchar`-CMA}}

\newcommand{\SPS}{{\mathsf{SPS}}}
\newcommand{\SPSSetup}{\mathsf{SPS.Setup}}
\newcommand{\SPSKeyGen}{\mathsf{SPS.KeyGen}}
\newcommand{\SPSSign}{\mathsf{SPS.Sign}}
\newcommand{\SPSVerify}{\mathsf{SPS.Verify}}

\newcommand{\MppDS}{M_{\ppDS}}
\newcommand{\KPW}{\mathsf{KPW}}
\newcommand{\PRSPS}{{\mathsf{PRSPS}}}
\newcommand{\Rand}{\mathsf{Rand}}

\newcommand{\RS}{{\mathsf{RS}}}
\newcommand{\RSO}{{\mathsf{RS}_{\mathsf{Ours}}}}

\newcommand{\RSSetup}{\mathsf{RS.Setup}}
\newcommand{\RSKeyGen}{\mathsf{RS.KeyGen}}
\newcommand{\RSSign}{\mathsf{RS.Sign}}
\newcommand{\RSRedact}{\mathsf{RS.Redact}}
\newcommand{\RSVerify}{\mathsf{RS.Verify}}
\newcommand{\ppRS}{\mathsf{pp}_{\RS}}
\newcommand{\pkRS}{\mathsf{pk}_{\RS}}
\newcommand{\skRS}{\mathsf{sk}_{\RS}}
\newcommand{\Mpp}{M_{\ppRS}}

\newcommand{\Uf}{\mathsf{Uf}}
\newcommand{\LR}{\mathsf{LoRredact}}

\newcommand{\SC}{\mathsf{SC}}
\newcommand{\SCSetup}{\mathsf{SC.Setup}}
\newcommand{\SCKeyGen}{\mathsf{SC.KGen}}
\newcommand{\SCCommit}{\mathsf{SC.Commit}}
\newcommand{\SCOpen}{\mathsf{SC.Open}}
\newcommand{\SCOpenSubset}{\mathsf{SC.OSubset}}
\newcommand{\SCVerifySubset}{\mathsf{SC.VSubset}}
\newcommand{\ppSC}{\mathsf{pp}_{\SC}}
\newcommand{\Spp}{S_{\ppSC}}
\newcommand{\ckSC}{\mathsf{ck}_{\SC}}
\newcommand{\Bind}{\mathsf{Bind}}
\newcommand{\Hide}{\mathsf{Hide}}
\newcommand{\state}{\mathsf{state}}
\newcommand{\Sound}{\mathsf{Sound}}

\newcommand{\OpenSubset}{\mathsf{OpenSubset}}
\newcommand{\FHS}{\mathsf{FHS}}

\begin{document}

\title{Redactable Signature with Compactness from Set-Commitment\thanks{This paper is appeared in IEICE Trans.Fundamentals \cite{TT21}.}}

\author{Masayuki Tezuka\textsuperscript{(\Letter)} \and Keisuke Tanaka}
\authorrunning{M. Tezuka et al.}
\institute{Tokyo Institute of Technology, Tokyo, Japan\\
\email{tezuka.m.ac@m.titech.ac.jp}}

\maketitle 
\pagestyle{plain}
\noindent
\makebox[\linewidth]{February 20, 2022}
             
\begin{abstract}
Redactable signature allows anyone to remove parts of a signed message without invalidating the signature.
The need to prove the validity of digital documents issued by governments is increasing. 
When governments disclose documents, they must remove private information concerning individuals.
Redactable signature is useful for such a situation.

However, in most redactable signature schemes, to remove parts of the signed message, we need pieces of information for each part we want to remove.
If a signed message consists of $\ell$ elements, the number of elements in an original signature is at least linear in $\ell$.

As far as we know, in some redactable signature schemes, the number of elements in an original signature is constant, regardless of the number of elements in a message to be signed. 
However, these constructions have drawbacks in that the use of the random oracle model or generic group model.

In this paper, we construct an efficient redactable signature to overcome these drawbacks.
Our redactable signature is obtained by combining set-commitment proposed in the recent work by Fuchsbauer et~al. (JoC 2019) and digital signatures.

\end{abstract}

%\setcounter{tocdepth}{2}
%\tableofcontents

\section{Introduction}
\subsection{Background}
Digital signature is an important cryptographic tool for data authentication. 
This allows a verifier to authenticate messages by verifying the signature.
By using digital signature, we can ensure that a message has not been modified since it was signed.
This property is useful for many scenarios.

However, in some scenarios, some limited modification of the signed message is desirable.
For example, we consider a situation where a citizen requests a secret signed document disclosure to the government.
To disclose the secret signed document, if privacy information is contained in the signed document, the government must remove this information from the signed document.
In a digital signature scheme, to ensure the validity of the modified document, the signer must resign this modified document.
If the original signer is not reachable anymore, or resigning the modified document produces too much overhead, it is not convenient.

Redactable signature is a useful cryptographic tool for the above situation.
This scheme allows anyone to remove parts of the message from the signed message and update its signature without a signing key.
We can check the validity of signed messages or submessages derived from signed messages.
Note that, in the context of redactable signature, an operation that removes some parts from a signed document is called ``redaction" and a message removed some parts of the signed message is called ``submessage".

The idea of redactable signature was introduced by Steinfeld, Bull, and Zheng \cite{SBZ01} as a content extraction signature.
This allows generating an extracted signature on selected portions of the signed original document while hiding removed parts of portions. Johnson, Molnar, Song, and Wagner \cite{JMSW02} proposed a redactable signature which is similar to a content extraction signature.
In addition, Miyazaki, Susaki, Iwamura, Matsumoto, Sasaki, and Yoshiura \cite{MSIMSY03} proposed the digital document sanitizing problem, and proposed the redactable signature scheme called SUMI-4.

Note that, in early studies of redactable signatures, the term ``sanitizing" indicates ``removing".
Later, Ateniese, Chou, de Medeiros, and Tsudik \cite{ACMT05} fromalized sanitizable signature.
They use the term ``sanitizing" to indicate "modifying".
That is, sanitizable signature allow some specific party to "rewrite" some message parts.
It is necessary to pay attention to distinguish whether the term ``sanizizable signature" indicates redactable signature or sanitaizable signature in the sense by Ateniese et al. \cite{ACMT05}.

Redactable signature has been studied for fundamental message data structures such as sets and lists.
Redactable signature has been extended to more complex data structures such that trees \cite{BBDFFKMOPPS10,CLX09,PSMP12,SPBPMtree12}, graphs \cite{KB13}, super-sets \cite{NTKK09}.

Security of redactable signature has been argued in many works.
Most works consider the following two security notions in common.
Unforgeability: An adversary cannot produce a signature for $M'$ except for any redacted version of an already signed one.
Privacy: Except for a signer and redactors, it is hard to derive information on redacted message parts when given a redacted message-signature pair.

Some works (e.g., \cite{BBDFFKMOPPS10,DPSS15,PS14,SPBPM12}) consider the security notion called transparency which strengthens the notion of privacy.
Transparency requires that it is hard to distinguish whether a signature $\sigma$ is an original signature or redacted ones.

Camenisch, Dubovitskaya, Haralambiev, and Kohlweiss \cite{CDHK15} proposed unlinkable redactable signature.
This signature satisfies unforgeability and unlinkability which is a variant security notion of privacy.
They used an unlinkable redactable signature scheme to construct an anonymous credential scheme \cite{Cha85}. 
Later, Sanders \cite{San20} also constructed an unlinkable redactable signature scheme to obtain an efficient anonymous credential scheme.

Currently, there are many studies for redactable signature and for its variants.
However, due to space limitations, we only mention principal researches related to our works.
See \cite{BPS17,DDHPST15} for a more comprehensive overview of studies for redactable signature and its variants.

\subsection{Motivation}\label{Moti}
In the use of a redactable signature, there is a problem we should consider.
Let us consider the following situation.
The government stores many secret signed original documents to a private cloud server.
For a disclosure request of a secret signed document from a citizen, the government officer retrieves the signed document from the private cloud server, remove privacy information, and disclose subdocument of the signed document.

When the government uses the private cloud server to store original signatures, there is a problem.
If the size of an original signature is too large, it takes too much time to upload the original signature to the private cloud server due to the limitation of internet communication bandwidth.
Unfortunately, many redactable signature schemes have a linear signature size in the number of elements in a message.
This makes it difficult to achieve quick uploading of an original signature to the private cloud server.

To overcome this problem, we requires that the size of an original signature is always constant regardless of the number of elements in a message.\footnote{In this work, the size of the signature is measured by the bit length of signature encoded in a bit string.}
Moreover, it is also desirable that the government officer can remove any parts from the original document.
How do we achieve these requirements?
This is a natural problem for a practical use of redactable signature.
Thus, we newly propose this problem as ``storing redactable signature problem".

To solve the ``storing redactable signature problem'', we require the ``compactness" for an original signature.
That is, the size of an original signature is always constant regardless of the number of elements in a message.\footnote{More precisely, we say that redactable signature satisfies compactness if the size of both an original signature and signature for a subdocument (redacted message) are always constant regardless of the number of elements in messages.}

Most redactable signature schemes with sets or lists seem hard to solve this problem.
Since the number of elements in an original signature is at least linear in the number of elements in an original message.

As far as we know, redactable signature schemes in \cite{ABCHSW12,CDHK15,San20} satisfy compactness.
However, these schemes have drawbacks.
The redactable signature scheme by Ahn, Boneh, Camenisch, Hohenberger, Shelat, and Waters \cite{ABCHSW12}\footnote{In \cite{ABCHSW12}, Ahn et al. proposed $\mathcal{P}$-homomorphic signature schemes where $\mathcal{P}$ is a predicate. This scheme allows anyone to derive a signature on the object $m'$ from a signature of $m$ as long as $\mathcal{P}(m, m')=1$ for the predicate $\mathcal{P}$. If we set $\mathcal{P}$ such that $\mathcal{P}(m, m') = 1$ if $m'$ is a subdocument of $m$ and $\mathcal{P}(m, m') = 0$ if $m'$ is not a subdocument of $m$, we can use $\mathcal{P}$-homomorphic signature schemes as redactable signature schemes.} uses the random oracle model (ROM) \cite{BR93}.
The redactable signature scheme by Sanders \cite{San20}, its security is guaranteed in the generic group model (GGM) \cite{Sho97}.
It is desirable to solve the ``storing redactable signature problem'' without the GGM or ROM.

As for the redactable signature scheme by Camenisch et al. \cite{CDHK15}, security of thier redactable signature scheme relies on the $\ell$-RootDH assumption \cite{CDHK15}, $(\ell + 1)$-BSDH assumption \cite{Goy07}, and the existentially unforgeable under chosen-message attacks (EUF-CMA) security for partial randomizable structure-preserving signatures (SPS) where $\ell$ is an upper bound for the number of elements in a message to be signed.
As far as we know, two partial randomizable SPS scheme exists. 
One is proposed by Abe, Fuchsbauer, Groth, Haralambiev, and Ohkubo \cite{AFGHO10}. 
The EUF-CMA security of this scheme is proven under the $q$-simultaneous flexible pairing ($q$-SFP) assumption \cite{AFGHO10} where $q$ is the number of signatures issued by the singer.
The other is proposed by Abe, Groth, Haralambiev, and Ohkubo \cite{AGHO11}.
This scheme has an optimal signature size. That is, a signature is composed of only $3$ group elements.
However, the security of this scheme is proven in GGM.
If we avoid using GGM and adapt the partial randomizable SPS by Abe et al. \cite{AFGHO10} to redactable signature scheme by Camenisch et al., the security of this redactable signature scheme relies on three $q$-type assumptions: {$\ell$}-RootDH; $(\ell + 1)$-BSDH; and $q$-SFP assumptions where $q$ is the number of original signatures issued by the singer.
These assumptions are not standard.
It is desirable to construct a redactable signature scheme whose security can be proven  with two or less q-type assumptions.

\subsection{Our Results}\label{OurResu}
In this paper, we give a new redactable signature scheme with compactness from a set-commitment scheme and a digital signature scheme.

A set-commitment scheme proposed by Fuchsbauer, Hanser, and Slamanig \cite{FHS19} allows us to commit to a set.
This supports ordinary opening and supports subsets opening.
Specifically, in a set-commitment scheme, we can commit to set $S$ and generate a commitment $C$ and its opening information $O$.
Moreover, from $(S, C, O)$, we can generate a witness $W$ for a subset $S' \subseteq S$.
By using $(S, C, O)$, we can verify that $S$ is committed to $C$.
Also, by using $(S', C, W)$, we can verify that $S'$ is a subset of $S$ which is committed to $C$.
Fuchsbauer et al. \cite{FHS19} constructed a set-commitment scheme under the $q$-co-discrete logarithm $(\qcoDL)$ assumption \cite{FHS19} and the $q$-co-generalized-strong-Diffie-Hellman $(\qcoGSDH)$ assumption \cite{FHS19}.
Moreover, they constructed attribute-based anonymous credentials by combining set-commitment and structure-preserving signatures on equivalence classes (SPS-EQ) works on the type 3 pairings.

Here, we briefly explain the idea of our redactable signature construction.
The property of set-commitment is similar to the property of redactable signature for set message structures.
Redactable signature with sets allows us to derive a submessage $M' \subseteq M$ from the signed message $M$.
The key idea is to combine the set-commitment scheme with redactable signature scheme.
The signature $\sigma$ on an original set-structured message $M$ is composed of $(C, \sigma_{\DS}, O)$ where $(C, O)$ is a set commitment and opening information pair computed by committing $M$, and $\sigma_{\DS}$ is a digital signature on $C$. 
Redaction from an original message $M$ to a submessage $M'$ is can be done by deriving a witness $W$ for $M'$ by using $(M, C, O)$. 
The redactable signature for the message $M'$ is composed of $(C, \sigma_{\DS}, W)$. 
See Section \ref{Ourconstreda} for our construction.

\begin{figure*}[htbp]
\begin{center}

\begingroup
\scalefont{0.8}
\begin{tabular}{llcccccccc}\hline

\begin{tabular}{l} Scheme \end{tabular} &Assumption & Mstr& $\pp$ + $\pk$ size &sig size& T & U & C & R \\
\hline
\hline
\multirow{2}{*}{
\begin{tabular}{l}
 MHI \\
 \S 3.2 in \cite{MHI06} 
 \end{tabular}
 }

 &BGLS-aggregate signature \cite{BGLS03} & \multirow{2}{*}{Set} & \multirow{2}{*}{$|\BG| + |\Hash| + |\G_2|$} & \multirow{2}{*}{\begin{tabular}{c}$(\ell+1)|\G_1| + |{\rm {DID}}| + 2\ell$ \end{tabular}} & \multirow{2}{*}{\begin{tabular}{c}$\times$ \end{tabular}}& \multirow{2}{*}{$\times$}&\multirow{2}{*}{$\times$}& \multirow{2}{*}{M}\\
 &based on co-CDH \cite{BLS01} and ROM&&&& &\\
 \hdashline

\multirow{2}{*}{
\begin{tabular}{l}  
SPBPM \\
\S 4.2 in \cite{SPBPM12}
\end{tabular}
}

&
BGLS-aggregate signature \cite{BGLS03} 
&\multirow{2}{*}{Set}& \multirow{2}{*}{$|\BG| + |\Hash| + |\G_2|$} & \multirow{2}{*}{$\frac{\ell(\ell+1)}{2}|\G_1| + |{\rm {DID}}| + \ell|r|$} & \multirow{2}{*}{$\checkmark$} & \multirow{2}{*}{$\times$} &\multirow{2}{*}{$\times$} & \multirow{2}{*}{M}\\
& based on co-CDH \cite{BLS01} and ROM&&&&\\
\hdashline

\multirow{2}{*}{
\begin{tabular}{l} 
ABCHSW \\
\S 4.2 in \cite{ABCHSW11full}
\end{tabular}
}

&Accumulator based on & \multirow{2}{*}{Set} & \multirow{2}{*}{$|N| + |\Hash| + |\mathbb{Z}_N|$} & \multirow{2}{*}{$|\mathbb{Z}_N|$} & \multirow{2}{*}{$\checkmark$} & \multirow{2}{*}{$\times$} & \multirow{2}{*}{\pmb{$\checkmark$}}& \multirow{2}{*}{M}\\
&RSA and {\bf ROM}&&&&&\\
\hdashline

\multirow{2}{*}{
\begin{tabular}{l}  
PS \\
\S 3 in \cite{PS14} 
\end{tabular}
}

&
Trapdoor accumulator based on
&\multirow{2}{*}{Set} & \multirow{2}{*}{$|N| + |\Hash| + |\mathbb{Z}_N|$} & \multirow{2}{*}{$(\ell + 1)|\mathbb{Z}_N|+ |{\rm {DID}}|$} & \multirow{2}{*}{$\checkmark$}& \multirow{2}{*}{$\times$}&\multirow{2}{*}{$\times$}& \multirow{2}{*}{M}\\
& GHR-signature \cite{GHR99} &&&\\
\hdashline

\multirow{3}{*}{
\begin{tabular}{l}
DPSS${}^{\dagger}$\\
\S 4 in \cite{DPSS15}
\end{tabular}
}

 &HW signature \cite{HW09} based on RSA & \multirow{3}{*}{Set}& \multirow{3}{*}{\tlef \begin{tabular}{c}$|\CH| + |\prfF| + |\prfK|$ \\ $+ |\prfout| + |N| + |\mathbb{Z}_N|$\end{tabular}\tlef } & \multirow{3}{*}{\begin{tabular}{c}$(\ell + 2) |\mathbb{Z}_N| + |\rHW|$ \\ $(3 |\mathbb{Z}_N| + |\rHW|)^{\dagger}$ \end{tabular}}& \multirow{3}{*}{$\checkmark$} & \multirow{3}{*}{$\times$} & \multirow{3}{*}{$\times^{\dagger}$} & \multirow{3}{*}{
M}\\
 & + unbounded accumulator \cite{DHS15}& & \\
 &~~based on strong-RSA &&&\\
 \hdashline

\multirow{4}{*}{
\begin{tabular}{l}
DPSS${}^{\dagger}$\\
\S 5 in \cite{DPSS15}
\end{tabular}
}
&
\hspace{-10pt}
\multirow{4}{*}{
\begin{tabular}{l}
HW signature \cite{HW09} based on RSA\\
+ unbounded accumulator \cite{DHS15}\\
~~based on strong-RSA
\end{tabular}
}

& \multirow{4}{*}{List} & \multirow{4}{*}{\tlef \begin{tabular}{c}$|\CH| + |\prfF| + |\prfK| $\\ $+ |\prfout| + |N| + |\mathbb{Z}_N| $ \end{tabular} \tlef } &\multirow{4}{*}{\tlef \begin{tabular}{c}$\left(1 +\frac{1}{2}\ell(\ell + 3)\right)|\mathbb{Z}_N|$\\ $+ \ell|r|+|\rHW|$ \\ $((2\ell + 1)|\mathbb{Z}_N| + |\rHW|)^{\dagger}$ \end{tabular} \tlef } & \multirow{4}{*}{$\checkmark$} & \multirow{4}{*}{$\times$} & \multirow{4}{*}{$\times^{\dagger}$} &\multirow{4}{*}{M}\\
 & & & \\
 & &&\\
 &&&\\
 \hdashline

\multirow{5}{*}{
\begin{tabular}{l}
CDHK${}^{\natural}$\\
\S 3.3 in \cite{CDHK15} 
\end{tabular}
}

&AFGHO-partial randomizable 
& \multirow{6}{*}{List} & \multirow{6}{*}{\tlef \begin{tabular}{c}$|\BG| + (\ell+15)|\G_1|$\\$+ (\ell+9)|\G_2|$ \end{tabular} \tlef } & \multirow{6}{*}{\begin{tabular}{c}$3|\G_1| + 5 |\G_2| + |\mathbb{Z}_p|$\\ $(18|\G_1| + 18|\G_2|)^{\natural}$\end{tabular}} & \multirow{6}{*}{$\times$} & \multirow{6}{*}{$\checkmark$} & \multirow{6}{*}{\pmb{$\checkmark$}} & \multirow{6}{*}{O}\\
 &SPS \cite{AFGHO10} based on $q$-SFP\\
 &+ vector-commitment \cite{CDHK15} based &&&\\
 & ~~on $\ell$-RootDH and $(\ell + 1)$-BSDH&&&\\
 & + GS extractable WI-PoK \cite{GS08}&&&\\
 & ~~ based on SXDH \\
 \hdashline

 \multirow{3}{*}{
\begin{tabular}{l}
 Sanders\\
 \S 4.2 in \cite{San20}
 \end{tabular}
 }

 &\multirow{3}{*}{{\bf GGM} (generic group model)} & \multirow{3}{*}{List} & \multirow{3}{*}{\begin{tabular}{c}$|\BG| + \left(1 +\frac{\ell(\ell+1)}{2}\right)|\G_1| + \ell|\G_2|$ \\ $\left( |\BG| + (\ell+1)|\G_1| + 
 \ell|\G_2| \right)^\diamondsuit$\end{tabular} } & \multirow{3}{*}{$2|\G_1|+2|\G_2|$} & \multirow{3}{*}{$\times$} & \multirow{3}{*}{$\checkmark$} & \multirow{3}{*}{$\pmb{\checkmark}$} &\multirow{3}{*}{O}\\
 & &&&\\
 \\
 \hdashline

\multirow{2}{*}{
\begin{tabular}{l}
$\RSO^{\flat}$ \\ Section \ref{Ourconstreda}
\end{tabular} 
}
& KPW-SPS \cite{KPW15} based on SXDH
&
\multirow{2}{*}{Set} & \multirow{2}{*}{$|\BG| + \ell|\G_1| + (\ell +4)|\G_2|$} &\multirow{2}{*}{\tlef \begin{tabular}{c}$8|\G_1| + |\G_2| +1$ \\$(8|\G_1| + |\G_2|)^{\flat}$\end{tabular} \tlef } & \multirow{2}{*}{$\times$}& \multirow{2}{*}{$\times$} & \multirow{2}{*}{$\pmb{\checkmark}$} & \multirow{2}{*}{O}\\
 & + set-commitment \cite{FHS19} based on &&&\\
 & ~~$\ellcoDL$ and $\ellcoGSDH$ &\multicolumn{5}{l} {\bf (avoid using ROM and GGM)}
\\
\hline
\end{tabular}\\
\endgroup
\end{center}

\caption{\small The comparison with major redactable signature schemes with privacy security.} 
$\ell$ is the number of elements of sets or lists to be signed.
The column ``Mstr" indicates the message data structure supported by the corresponding scheme.
The column ``$\pp$ + $\pk$ size" represents the sum of the public parameters bit length and a public key bit length.
The column ``sig size" represents the signature bit length.
$|\BG|$, $|\Hash|$, $|\CH|$ indicate the bit length of description for a bilinear group $\BG$, a hash function $\Hash$ and a chameleon hash function $\CH$, respectively.
$|\G_1|$ and $|\G_2|$ indicate the bit length of an element in $\G_1$ and $\G_2$, respectively.
For an integer $N = pq$ where $p$ and $q$ are distinct primes, $|\mathbb{Z}_N|$ indicates the bit length of an element in $\mathbb{Z}_N$.
For a pseudorandom function $\prfF$, $|\prfF|$, $|\prfK|$, and $|\prfout|$ represent the bit length of description for the pseudorandom function $\prfF$, the bit length of a key for $\prfF$, and the output bit length of $\prfF$, respectively.
$|\crs|$ and $|\pi|$ are the bit length of the common reference strings $\crs$ and the proof $\pi$ for the extractable WI-PoK system.
$\DID$ represents a document ID. In the SPBPM and PS schemes, $\DID$ is called ``tag". $|\DID|$ represents the bit length of $\DID$ whose length is polynomial in security parameters.
$|\rHW|$ denotes the bit length of a random string $\rHW$ which is originated from HW signature.
$|r|$ denotes the bit length of randomness $r$.
Both $|\rHW|$ and $|r|$ are is polynomial in security parameter.
The columns ``T", ``U" and ``C" represent transparency, unlinkability and compactness respectively.
The checkmark $\checkmark$ represents that the scheme satisfies the security of the corresponding column.
In the column ``R", M represents that the scheme supports multiple-time redaction and O represents that the scheme only supports one-time reduction.
\protect
${}^{\dagger}$ The DPSS schemes are constructed from unbounded accumulators and EUF-CMA secure signatures in a black-box way. For this reason, public parameters, public key, and signature size of the DPSS schemes are described only asymptotically.
To compare with other redactable signature schemes, we apply the short RSA signature scheme by Hohenberger and Water \cite{HW09} in the standard model and unbounded accumulator \cite{DHS15} based on the strong-RSA assumption to the DPSS schemes.
The DPSS schemes can issue two types of signatures. 
One type is not short but redactable.  
The other type is short but not redactable.
The obvious approach to compactness is to generate a later type signature for the original message. 
However, this approach does not achieve redactability.
\protect
${}^{\natural}$ The CDHK scheme is constructed from partial randomizable structure-preserving signatures, vector-commitments, and witness-indistinguishable proof-of-knowledge (extractable WI-PoK) system in a black-box way.
For this reason, public parameters, public key, and signature size of the CDHK scheme are analyzed only asymptotically.
To compare with other redactable signature schemes, we apply the partial randomizable structure-preserving signature scheme \cite{AFGHO10} based on the $q$-SFP assumption, vector-commitment based on $\ell$-RootDH and $(\ell + 1)$-BSDH assumption to the CDHK scheme where $q$ is the number of original signatures issued by the singer, and GS extractable WI-PoK \cite{GS08} based on the SXDH assumption.
In this instantiation, the size of a redacted signature is longer than that of an original signature.
\protect
${}^\diamondsuit$ In Sanders scheme, if we verify the validity of a signature, we only use $O(\ell)$ elements in the public key.
$O(\ell^2)$ elements in the public key are needed to support redaction operations. 
\protect
$^{\flat}$ Our scheme $\RSO$ is constructed from digital signatures and set-commitments.
To compare with other redactable signature schemes, we apply the structure-preserving signature by Kiltz et al. \cite{KPW15} and the set-commitment by Fuchsbauer et al. \cite{FHS19} to $\RSO$.
In this instantiation, the size of a redacted signature is shorter than that of an original signature.%\protect
\label{PBSinst}

\end{figure*}
Our redactable signature scheme for sets is constructed from set-commitment and a digital signature.
To compare our redactable signature scheme with other redactable signature schemes, we consider the concrete instantiation for our scheme. We instantiate a redactable signature scheme by adopting the set-commitment scheme by Fuchsbauer et al. and the structure-preserving signature by Kiltz, Pan, and Wee \cite{KPW15}. 
We explain the reason why we adopt the structure-preserving signature by Kiltz et al.
Firstly, both the set-commitment scheme by Fuchsbauer et al. and the structure-preserving signature by Kiltz et al. work on type 3 pairings.
Secondly, in the set-commitment scheme proposed by Fuchsbauer et al., the commitment $C$ belongs to $\G_1$.
We need a signature scheme that supports $\G_1$ element signing.
The structure-preserving signature by Kiltz et al. supports $\G_1$ element signing.
Finally, the structure-preserving signature by Kiltz et al. is efficient and its security is proven without GGM or ROM.
Our instantiated redactable scheme can be proven under the $\ellcoDL$, $\ellcoGSDH$, and $\SXDH$ assumption \cite{BGMM05} where $\ell$ is an upper bound for the number of elements in a message to be signed.

We summarize redactable signature scheme with compactness and major redactable signature schemes for sets or lists in Fig. \ref{PBSinst}.
Our redactable signature scheme is a better solution for the ``storing redactable signature problem" than other redactable signatures schemes with compactness \cite{ABCHSW12,San20} in that our scheme does not use the GGM or ROM.
The redactable signature scheme by Camenisch et al.  instantiated by the partial randomizable SPS by Abe et al. \cite{AFGHO10} relies on three $q$-type assumption. 
Compared with this redactable signature scheme, our scheme is milder in that our scheme can be proven with two $q$-type assumptions to two (the $l$-co-DL and $l$-co-GSDH assumptions).
Moreover, in the security redactable signature by Camenisch et al., the parameter $q$ of the $q$-SFP assumption depends on the number of signatures issued by the signer.
By contrast, all assumptions (the $l$-co-DL, $l$-co-GSDH, and SXDH assumptions) we need to prove the security of our scheme are independent of the number of signatures issued by the signer.

Furthermore, in the redactable signature scheme by Camenisch et al., to generate a redacted version of signature, we must prove pairing equations by using a WI-PoK proof system.
For this reason, this causes somewhat large signature size.
We estimate the size of redacted version of a signature of their scheme in Fig. \ref{PBSinst} in the case of adapting the Groth-Sahai extractable WI-PoK system \cite{GS08} based on the SXDH assumption.
By comparing instantiations of our redactable signature scheme and that of scheme by Camenisch et al. in Fig. \ref{PBSinst},
our scheme has advantage  with the concrete instantiation of their method, our scheme has shorter redacted signature size.

Our redactable signature scheme is similar to the redactable signature scheme by Camenisch et al. \cite{CDHK15}. 
In their redactable signature scheme, the signature $\sigma$ on an original vector-structured (list-structured) message $M$ is composed of $(C, \sigma_{\PRSPS}, O)$ where $(C, O)$ is a vector commitment and opening information pair computed by committing $M$, and $\sigma_{\PRSPS}$ is partial randomizable digital signature on $C$.
There is a difference in deriving the redactable signature.
In their redactable signature scheme, redaction from an original message $M$ to a submessage $M'$ is proceed as follows.
First, we derive a witness $W$ for $M'$, randomize $\sigma_{\PRSPS}$ to $\sigma'_{\PRSPS}$.
Then, we parse $\sigma'_{\PRSPS}$ as fixed elements $\sigma'^{\Fix}_{\PRSPS}$ and randomized elements $\sigma'^{\Rand}_{\PRSPS}$ and generates a proof $\pi$ for the knowledge of $(C, W, \sigma'^{\Fix}_{\PRSPS})$ by using witness-indistinguishable proof-of-knowledge (WI-PoK) system.
The redactable signature for the message $M'$ is composed of $(\pi, \sigma'^{\Rand})$. 
The main difference between our scheme and scheme by Camenisch et al. is the use of witness-indistinguishable proof-of-knowledge (WI-PoK).

We briefly explain the reason why their redactable signature scheme uses WI-PoK.
Their redactable signature scheme was constructed to satisfy unlinkability.
Unlinkability requires that it should be hard to link back a redacted signature to its original signature.
If the original signature and its redacted signature share a common (fixed) part, it is easy to link back from the redacted signature to its original signature. 
To hide common (fixed) parts in the redacted signature, they used a WI-PoK proof.

Although our scheme does not have transparency and unlinkability, our scheme makes sense in the following points.
Non-transparent redactable signature has the drawback that an adversary can recover removed parts by collecting multiple submessages for an original signed message.
That is, by comparing multiple submessages with different removed parts for the same original message, the adversary recovers removed parts of the original message.
This attack can be avoided by restricting the number of redacted signatures for each original message to one.
For example, we consider a situation where the government issues a subdocument only once for each signed document.
In this situation, the adversary cannot obtain multiple redacted subdocuments for an original signed document, we can avoid this attack.
Moreover, many redactable signatures without transparency \cite{HHHHMSTY08,IIKO11,IKOST08,IKOST09,MLWW17,MHI06,MHI08,MIMSYTI05} were constructed for real scenarios where non-transparency is desirable.

Unlinkability for a redactable signature scheme is useful to construct anonymous credential schemes.
However, to solve the ``storing redactable signature problem", unlinkability security is too strong.
We require that a redactable signature scheme satisfies only privacy, because it is sufficient to solve ``storing redactable signature problem".
Hence, our redactable signature scheme is suitable for solving the``storing redactable signature problem" in the situation where the number of redacted signatures for each original message is restricted to one.

\subsection{Related works}
We present several signatures that allow editing a signed message.
\begin{itemize}
\item Append-only signature \cite{KMPR05}: Kiltz, Mityagin, Panjwani, and Raghavan \cite{KMPR05} introduced the notion of append-only signature.
In this signature, we can only publicly append message blocks to a signed message and update the signature correspondingly.

\item Sanitizable signature \cite{ACMT05}: Ateniese, Chou, de Medeiros, and Tsudik~\cite{ACMT05} introduced the notion of sanitizable signature.
In this signature, a signer selects a sanitizer who can modify the signed message and generate a signature.
The sanitizer can modify some parts of message blocks of the signed document, but he or she cannot remove message blocks.
In the redactable signature, anyone can redact parts of the signed message without the secret key. 
However, in the sanitizable signature scheme, each sanitizer has the sanitizer's secret key and the sanitizer designated by the signer can sanitize parts of the message using own sanitizer's secret key.

\item Protean signature \cite{KPSS18}: Krenn, P{\"{o}}hls, Samelin, and Slamanig \cite{KPSS18} introduced the notion of protean signature.
This signature allows removing and editing some parts of message blocks. 
They give the construction of the protean signature scheme from a sanitizable signature scheme and a redactable signature scheme in the black-box way.
\end{itemize}

\subsection{Road Map}
In Section \ref{Prelimin}, we introduce notations and recall digital signature. 
In Section \ref{SCSch}, we review set-commitment and its security notions by Fuchsbauer et al. \cite{FHS19}. 
In Section \ref{RSintro}, we review redactable signature and its security notions.
In Section \ref{Ourconstreda}, we give a construction of the redactable signature scheme and its security analysis.
In \ref{RSOunfappen}, we provide a missing security proof for our scheme. In \ref{SetComFHS}, we review bilinear groups, the structure-preserving signature by Kiltz et al. \cite{KPW15}, and the set-commitment construction by Fuchsbauer et al. \cite{FHS19}.

\section{Preliminaries}\label{Prelimin}
\subsection{Notations}
Let $1^{\lambda}$ be the security parameter. 
A function $f(\lambda)$ is negligible in $\lambda$ if $f(\lambda)$ tends to $0$ faster than $\frac{1}{\lambda^c}$ for every constant $c > 0$.
PPT stands for probabilistic polynomial time.
For an integer $n$, $[n]$ denotes the set $\{1,\dots, n\}$.
For a finite set $S$, $s \xleftarrow{\$} S$ denotes choosing an element $s$ from $S$ uniformly at random and $\#S$ denotes the number of elements in $S$. 
For a group $\G$, we define $\G^* := \G \backslash \{1_{\G}\}$ where $1_{\G}$ is the identity element of the group $\G$. 
For an algorithm $\A$, 
$y \leftarrow \A(x)$ denotes that the algorithm $\A$ outputs $y$ on input $x$.

\subsection{Digital Signature}\label{DSdef}
\begin{definition}[Digital Signature Scheme]
A digital signature scheme $\DS$ is composed of following four algorithms $(\DSSetup, \DSKeyGen, \DSSign, \DSVerify)$. 
$\DSSetup(1^{\lambda})$ takes security parameters and generates public parameters $\ppDS$ which defines the message space $\MppDS$. 
$\DSKeyGen(\ppDS)$ takes public parameters $\ppDS$, return a public key $\pkDS$ and a signing key $\skDS$. 
$\DSSign(\ppDS, \skDS, m)$ takes public parameters $\ppDS$, a signing key $\skDS$, and a message $m \in \MppDS$, return a signature $\sigma$.
$\DSVerify(\ppDS, \pkDS, m, \sigma)$ takes public parameters $\ppDS$, a public key $\pkDS$, a message $m \in \MppDS$, and a signature $\sigma$, return $1$ or $0$.
\end{definition}

For $\DS$, we require the following correctness. 
\begin{itemize}
\item{Correctness:} 
A digital signature scheme $\DS$ is correct if for all $\lambda \in \mathbb{N}$, $\ppDS \leftarrow \DSSetup(1^{\lambda})$, for all $m \in \MppDS$, $(\pkDS, \skDS) \leftarrow \DSKeyGen(\ppDS)$, $\sigma \leftarrow \DSSign(\ppDS, \skDS, m)$, then $\DSVerify(\ppDS, \pkDS, m, \sigma) = 1$ holds.
\end{itemize}

\begin{definition}[EUF-CMA]\label{dsufdef}
Existentially unforgeable under chosen-message attacks (EUF-CMA) security for a digital signature scheme $\DS$ is defined by the following unforgeability game between a challenger and an adversary $\A$.

\begin{itemize}
\item The challenger computes $\ppDS \leftarrow \DSSetup(1^{\lambda})$, $(\pkDS, \skDS) \leftarrow \DSKeyGen(\ppDS)$ initializes $Q \leftarrow \{\}$, and sends $(\ppDS, \pkDS)$ to $\A$. 
\item $\A$ is given access to a signing oracle $\mathcal{O}^{\Sign}(\cdot)$.
Given an input $m$, $\mathcal{O}^{\Sign}$ computes $\sigma \leftarrow \DSSign(\skDS, m)$, update $Q \leftarrow Q \cup \{m\}$ and returns $\sigma$ to $\A$.
\item Finally, $\A$ outputs a forgery $(m^*, \sigma^*)$.
\end{itemize}

$\DS$ is EUF-CMA secure if for all $\lambda \in \mathbb{N}$ and all PPT adversaries $\A$,
the advantage $\Adv^{\sfEUFCMA}_{\DS, \A}:= \Pr \left [ 
\DSVerify(\ppDS, \pkDS, m^*, \sigma^*)=1 \land m^* \notin Q  
\right]$ is negligible in $\lambda$.
\end{definition}

\section{Set-Commitment}\label{SCSch}
Fuchsbauer et al. \cite{FHS19} proposed set-commitment which allows us to commit to a set.
This scheme supports ordinary opening and subsets opening.
In particular, we can commit a set $S$ and generate a commitment $C$ and its opening information $O$.
Moreover, from $(S, C, O)$, we can generate a witness $W$ of a subset $S' \subseteq S$.
By using $(S, C, O)$, we can verify that $S$ is committed to $C$.
Also, by using $(S', C, W)$, we can verify that $S'$ is a subset of $S$ which is committed to $C$.
Now, we review the definition for set-commitment schemes.

\begin{definition}[Set-Commitment Scheme \cite{FHS19}]\label{SCSsyntax}
Let $\ell$ be a polynomial in $\lambda$.
A set-commitment scheme $\SC$ is a tuple of algorithms $(\SCSetup, \allowbreak \SCKeyGen, \SCCommit, \SCOpen, \allowbreak \SCOpenSubset, \allowbreak \SCVerifySubset)$.\footnote{In the syntax by Fuchsbauer et al. \cite{FHS19}, a set commitment scheme consists of five algorithms $(\SCSetup, \allowbreak \SCCommit, \SCOpen, \SCOpenSubset, \SCVerifySubset)$. In this work, we divide $\SCSetup$ in \cite{FHS19} into two algorithms $\SCSetup$ and $\SCKeyGen$ for convenience in constructing the redactable signature scheme.}
\begin{itemize}
\item $\SCSetup(1^{\lambda}):$ Given a security parameter $\lambda$, return public parameters $\ppSC$ which defines the message space $\Spp$.

\item $\SCKeyGen(\ppSC, 1^{\ell}):$ Given public parameters $\ppSC$ and an upper bound $\ell$ for the number of elements in committed sets, return a commitment key $\ckSC$.
$\ckSC$ supports committing for a non-empty set containing at most $\ell$ elements.

\item $\SCCommit(\ppSC, \ckSC, S):$ Given public parameters $\ppSC$, a commitment key $\ckSC$ and a non-empty set $S \subseteq \Spp$, return a commitment and opening information pair $(C, O)$ or $\bot$.
 
\item $\SCOpen(\ppSC, \ckSC, C, S, O):$ Given public parameters $\ppSC$, a commitment key $\ckSC$, a commitment $C$, a non-empty set $S \subseteq \Spp$, and opening information $O$, return $1$ (Valid) or $0$ (Invalid).

\item $\SCOpenSubset(\ppSC, \ckSC, C, S, O, S'):$ Given public parameters $\ppSC$, a commitment key $\ckSC$, a commitment $C$, a non-empty set $S \subseteq \Spp$, opening information $O$, and a non-empty subset $S' \subseteq S$, return a witness $W$ or~$\bot$.

\item $\SCVerifySubset(\ppSC, \ckSC, C, S', W):$ Given public parameters $\ppSC$, a commitment key $\ckSC$, a commitment $C$, a non-empty set $S' \subseteq \Spp$, and a witness $W$, return $1$ (Valid) or $0$ (Invalid).
\end{itemize}
\end{definition}

For $\SC$, we require the following correctness and compactness.
\begin{itemize}
\item{Correctness:} 
A set-commitment scheme $\SC$ is correct if for all $\lambda \in \mathbb{N}$, for all $\ell(\lambda)>0$, $\ppSC \leftarrow \SCSetup(1^{\lambda})$, for all non-empty set $S \subseteq \Spp$ where $\#S \leq \ell$, and for all non-empty subset $S' \subseteq S$, $\ckSC \leftarrow \SCKeyGen(\ppSC, 1^{\ell})$, $(C, O) \leftarrow \SCCommit(\ppSC, \ckSC, \allowbreak S)$, $W \leftarrow \SCOpenSubset(\ppSC, \allowbreak \ckSC, C, S, O, S')$, then followings holds.
\begin{equation*}
\begin{split}
\SCOpen(&\ppSC, \ckSC, C, S, O) = 1 \land \SCVerifySubset(\ppSC, \ckSC, C, S', W) = 1
\end{split}
\end{equation*}

\item{Compactness:}
A set-commitment scheme $\SC$ satisfies compactness if for all $\lambda \in \mathbb{N}$, for all $\ell(\lambda)>0$, $\ppSC \leftarrow \SCSetup(1^{\lambda})$, for all non-empty set $S \subseteq \Spp$ where $\#S \leq [\ell]$, and for all non-empty subset $S' \subseteq S$, $\ckSC \leftarrow \SCKeyGen(\ppSC, 1^{\ell})$, $(C, O) \leftarrow \SCCommit(\ppSC, \ckSC, \allowbreak S)$, $W \leftarrow \SCOpenSubset(\ppSC, \allowbreak \ckSC, C, S, O, S')$, the bit length of $C$, $O$, and $W$ are independent of $\ell$, $\#S$, and $\#S'$.
\end{itemize}

We review security notions for set-commitment.

\begin{definition}[Binding \cite{FHS19}]
A set-commitment scheme $\SC$ is computationally binding if for all $\lambda \in \mathbb{N}$, $\ell(\lambda)>0$, and all PPT adversaries $\A$, the following advantage 
\begin{equation*}
\begin{split}
&\Adv^{\Bind}_{\SC, \A}:= \Pr\left[ 
\begin{split}
&\SCOpen(\ppSC, \ckSC, C, S, O) = 1\\
&\land \SCOpen(\ppSC, \ckSC, C, S^{*}. O^{*}) = 1 \\
&\land S \neq S^{*}
\end{split}\middle| 
\begin{split}
&\ppSC \leftarrow \SCSetup(1^{\lambda})\\
&\ckSC \leftarrow\SCKeyGen(\ppSC, 1^{\ell})\\
&(C, S, O, S^{*}, O^{*}) \leftarrow \A(\ppSC,\ckSC)
\end{split}\right]
\end{split}
\end{equation*}
is negligible in $\lambda$.
\end{definition}

\begin{definition}[Subset-Soundness \cite{FHS19}]
A set-commitment scheme $\SC$ is subset-sound if for all $\lambda \in \mathbb{N}$, $\ell(\lambda)>0$, and all PPT adversaries $\A$, the following advantage 
\begin{equation*}
\begin{split}
&\Adv^{\Sound}_{\SC, \A}:=
 \Pr\left[ 
\begin{split}
&\SCOpen(\ppSC, \ckSC, C, S, O) = 1\\
&\land \SCVerifySubset(\ppSC, \ckSC,  C, S', W) = 1\\
&\land S' \nsubseteq S
\end{split}\middle| 
\begin{split}
&\ppSC \leftarrow \SCSetup(1^{\lambda})\\
&\ckSC \leftarrow \SCKeyGen(\ppSC, 1^{\ell})\\
&(C, S, O, S', W) \leftarrow \A(\ppSC,\ckSC)
\end{split}\right]
\end{split}
\end{equation*}
is negligible in $\lambda$.
\end{definition}

\begin{definition}[Hiding \cite{FHS19}]\label{SChiddef}
Hiding for a set-commitment scheme $\SC$ is defined by the following hiding game between a challenger and an adversary $\A$.
\begin{itemize}
\item The challenger chooses $b \xleftarrow{\$} \{0, 1\}$, computes $\ppSC \leftarrow \SCSetup(1^{\lambda})$ and $\ckSC \leftarrow \SCKeyGen(\ppSC, 1^{\ell})$.
Then, the challenger sends $(\ppSC, \ckSC)$ to $\A$. 
\item $\A$ sends a challenge $(S_{0}, S_{1}, \state)$ to the challenger.
\item The challenger computes $(C, O) \leftarrow \SCCommit(\ppSC, \allowbreak \ckSC, S_b)$.
Then, the challenger sends $(C, \state)$ to $\A$. 
\item $\A$ is given access to an open-subset oracle $\mathcal{O}^{\OpenSubset}(\cdot)$.
Given an input $S$, $\mathcal{O}^{\OpenSubset}$ computes $W \leftarrow \SCOpenSubset(\ppSC, \ckSC, C, S, O, (S \cap S_0 \cap S_1))$ and returns $W$ to $\A$.
\item Finally, $\A$ outputs a guess $b^{*}$
\end{itemize}

$\SC$ is computationally hiding if for all $\lambda \in \mathbb{N}$, all $\ell(\lambda)>0$, and all PPT adversaries $\A$,
the advantage $\Adv^{\Hide}_{\RS, \A}:= \left| \Pr \left [b^* =b \right] - \frac{1}{2}\right|$ is negligible in $\lambda$.
\end{definition}
We say $\SC$ is perfectly hiding if $\Adv^{\Hide}_{\SC, \A} = 0$ holds for all $\lambda \in \mathbb{N}$, $\ell(\lambda)>0$, and all PPT adversaries $\A$.

Fuchsbauer et al. \cite{FHS19} gave a set-commitment scheme which satisfies correctness, compactness, binding, subset-soundness, and hiding.

\section{Redactable Signature}\label{RSintro}

We review the definition of a redactable signature scheme and its security notions.
Our redactable signature scheme support sets signing.
We refer to the syntax of a redactable signature by Sanders \cite{San20}.
However, the syntax of the redactable signature scheme by Sanders is dedicated to redactable signature schemes for lists signing.
We tailor this syntax for the redactable signature scheme with sets.

\begin{definition}[Redactable Signature Scheme]
Let $\ell$ be a polynomial in $\lambda$.
A redactable signature scheme $\RS$ is composed of following five algorithms $(\RSSetup, \RSKeyGen, \allowbreak \RSSign, \RSRedact, \allowbreak \RSVerify)$. 

\begin{itemize}
\item $\RSSetup(1^{\lambda}):$ Given a security parameter $\lambda$, return public parameters $\ppRS$ which defines the message space $\Mpp$.

\item $\RSKeyGen(\ppRS, 1^{\ell}):$ Given public parameters $\ppRS$ and an upper bound $\ell$ for the number of elements in sets to be signed, return a public key $\pkRS$ and a signing key $\skRS$.
$\skRS$ supports signing for a non-empty set containing at most $\ell$ elements.

\item $\RSSign(\ppRS, \skRS, M):$ Given public parameters $\ppRS$, a signing key $\skRS$, and non-empty set $M \subseteq \Mpp$, return a signature $\sigma$ on the set $M$.
 
\item $\RSRedact(\ppRS, \pkRS, M, \sigma, M'):$ Given public parameters $\ppRS$, a public key $\pkRS$, a non-empty set $M \subseteq \Mpp$, a signature $\sigma$, and a non-empty subset $M' \subseteq M$, return a signature $\sigma'$ on the subset $M'$ or $\bot$.

\item $\RSVerify(\ppRS, \pkRS, M, \sigma):$ Given public parameters $\ppRS$, a public key $\pkRS$, a non-empty set $M \subseteq \Mpp$, and a signature $\sigma$, return $1$ (Valid) or $0$ (Invalid).
\end{itemize}
\end{definition}

For $\RS$, we require the following correctness and compactness.

\begin{itemize}
\item {Correctness:} 
A redactable signature scheme $\RS$ is correct if for all $\lambda \in \mathbb{N}$, for all $\ell(\lambda)>0$, $\ppRS \leftarrow \RSSetup(1^{\lambda})$, for all non-empty $M \subseteq \Mpp$ where $\#M\leq \ell$, and for all non-empty subset $M' \subseteq M$, $(\pkRS, \skRS) \leftarrow \RSKeyGen(\ppRS, 1^{\ell})$, $\sigma \leftarrow \RSSign(\ppRS, \skRS, M)$, $\sigma' \leftarrow \RSRedact(\ppRS, \pkRS, \allowbreak M, \sigma, M')$, then $\RSVerify(\ppRS, \allowbreak \pkRS, \allowbreak M, \sigma) = 1$ and $\RSVerify(\ppRS, \allowbreak \pkRS, \allowbreak M', \allowbreak \sigma') =1$ hold.

\item{Compactness:}
A redactable signature scheme $\RS$ satisfies compactness if for all $\lambda \in \mathbb{N}$, for all $\ell(\lambda)>0$, $\ppRS \leftarrow \RSSetup(1^{\lambda})$, for all non-empty $M \subseteq \Mpp$ where $\#M\leq \ell$, and for all non-empty subset $M' \subseteq M$, $(\pkRS, \skRS) \leftarrow \RSKeyGen(\ppRS, 1^{\ell})$, $\sigma \leftarrow \RSSign(\ppRS, \skRS, M)$, $\sigma' \leftarrow \RSRedact(\ppRS, \pkRS, \allowbreak M, \sigma, M')$, the bit length of both $\sigma$ and $\sigma'$ are independent of $\ell$, $\#M$, and $\#M'$.
\end{itemize}

We review unforgeability and privacy for redactable signature.
These security notions were formalized by Brzuska et al. \cite{BBDFFKMOPPS10} for redactable signature for tree message structures.
Later, these security notions were extended to redactable signature for arbitrary data structures by Derler et al. \cite{DPSS15}.

Unforgeability requires that without a signing key $\sk$, it should be infeasible to compute a valid signature $\sigma$ on $m'$ except to redact a signed message $(m, \sigma)$.

\begin{definition}[Unforgeability]\label{ufdef}
Unforgeability for a redactable signature scheme $\RS$ is defined by the following unforgeability game between a challenger and an adversary $\A$.

\begin{itemize}
\item The challenger computes $\ppRS \leftarrow \RSSetup(1^{\lambda})$, $(\pkRS, \skRS) \leftarrow \RSKeyGen(\ppRS, 1^{\ell})$ initializes $Q \leftarrow \{\}$, and sends $(\ppRS,\pkRS)$ to $\A$. 
\item $\A$ is given access to a signing oracle $\mathcal{O}^{\Sign}(\cdot)$.
Given an input $M$, $\mathcal{O}^{\Sign}$ computes $\sigma \leftarrow \RSSign(\ppRS, \skRS, M)$, update $Q \leftarrow Q \cup \{M\}$ and returns $\sigma$ to~$\A$.
\item Finally, $\A$ outputs a forgery $(M^*, \sigma^*)$.
\end{itemize}

$\RS$ is unforgeable if for all $\lambda \in \mathbb{N}$, $\ell(\lambda)>0$, and all PPT adversaries $\A$,
the following advantage 
\begin{equation*}
\begin{split}
&\Adv^{\Uf}_{\RS, \A} := \Pr \left [ 
\begin{split}&\RSVerify(\ppRS, \pkRS, M^*, \sigma^*)=1 \land \not\exists M \in Q : \ M^* \subseteq M 
\end{split}
\right]
\end{split}
\end{equation*}
is negligible in $\lambda$.
\end{definition}

Privacy requires that except for a signer and a redactor, it is infeasible to derive information on redacted message parts when given a redacted message-signature pair.
\begin{definition}[Privacy]
Privacy for a redactable signature scheme $\RS$ is defined by the following unforgeability game between a challenger and an adversary~$\A$.
\begin{enumerate}
\item The challenger computes $\ppRS \leftarrow \RSSetup(1^{\lambda})$, $(\pkRS, \skRS) \leftarrow \RSKeyGen(\ppRS, 1^{\ell})$, chooses $b \xleftarrow{\$} \{0, 1\}$, and sends $(\ppRS,\pkRS)$ to $\A$. 
\item $\A$ is given access to a signing oracle $\mathcal{O}^{\Sign}(\cdot)$.
Given an input $M$, $\mathcal{O}^{\Sign}$ computes $\sigma \leftarrow \RSSign(\ppRS, \skRS, M)$ and returns $\sigma$ to $\A$.

\item $\A$ is also given access to a left-or-right redact oracle $\mathcal{O}^{\LR}(\cdot, \cdot, \cdot)$. 
Given an input $(M_{0}, M_{1}, M')$, $\mathcal{O}^{\LR}$ works as follows:
\begin{enumerate}
\item If $M' \nsubseteq (M_{0} \cap M_{1})$, return $\bot$.
\item Compute $\sigma_{b} \leftarrow \Sign(\ppRS, \skRS, M_{b})$, $\sigma'_{b} \leftarrow \RSRedact(\ppRS, \pkRS, M_{b}, \sigma_{b}, M')$.
\item Return $\sigma'_{b}$.
\end{enumerate}
\item Finally, $\A$ outputs $b^*$.
\end{enumerate}

$\RS$ is private if for all $\lambda \in \mathbb{N}$, $\ell(\lambda)>0$, and all PPT adversaries $\A$,
the advantage $ \Adv^{\Priv}_{\RS, \A}:=  \left|\Pr \left [ b^* = b \right] -\frac{1}{2} \right|$ is negligible in $\lambda$.
\end{definition}

\section{Our Redactable Signature Scheme}\label{Ourconstreda}
In this section, we give a construction of a redactable signature scheme with compactness without the  GGM or ROM.
Then, we give security analysis for our redactable signature scheme.

\subsection{Our Construction} 
Before describing our construction, we give an intuition for our construction.
We can observe that a redactable signature scheme and a set-commitment scheme have similar properties.
In a redactable signature scheme, we can remove parts of a signed message without invalidating the signature.
That is, we can generate signatures for a subset of messages from the original signed document.
A set-commitment scheme has a similar flavor that we can generate a valid witness for subset opening for the committed sets.  

The key idea of our construction is combining set-commitment with digital signature. 
Let $(C, O)$ be a set commitment and opening information for set $M$.
A public key $\pkRS$ (resp., secret key $\skRS$) for our redactable signature scheme consists of $(\pkDS, \ckSC)$ (resp., $(\pkDS, \ckSC, \skDS)$) where $(\pkDS, \skDS)$ is a public key and signing key pair of the digital signature scheme and $\ckSC$ is a commitment key of the set commitment scheme.
A signature $\sigma$ for a message $M$ is composed of $(C, \sigma_{C}, O)$ where $(C, O)$ be a set commitment and opening information for set $M$, and $\sigma_{C}$ is a signature for a commitment $C$ generated by the digital signature scheme.
Redaction can be done by the following procedure.
Let $(M, \sigma = (C, \sigma_{C}, O))$ be a pair of an original message and signature.
To derive a signature $\sigma'$ for a subset message $M' \subset M$ from $(M, \sigma = (C, \sigma_{C}, O))$, we only change opening information $O$ to a witness $W$ for opening $M'$.
This can be done by the property of the set-commitment scheme.
A derived signature $\sigma'$ for $M'$ is formed as $\sigma' = (C, \sigma_{C}, W)$.
Thus, we can construct the redactable signature scheme from a digital signature scheme and a set-commitment scheme.

Let $\DS = (\DSSetup, \DSKeyGen, \DSSign, \DSVerify)$ be a digital signature scheme and $\SC = (\SCSetup, \allowbreak \SCKeyGen, \SCCommit, \SCOpen,  \allowbreak \SCOpenSubset, \allowbreak \SCVerifySubset)$ a set-commitment scheme.
The construction of our redactable signature scheme $\RSO$ is given in Fig.~\ref{RSOframe}.

\begin{figure}[h]
\centering
\begin{tabular}{|l|}
\hline
Algorithm $\RSSetup(1^{\lambda}):$\\
~~~$\ppSC \leftarrow \SCSetup(1^{\lambda}, 1^{\ell})$,  $\ppDS \leftarrow \DSSetup(1^{\lambda})$, $\ppRS \leftarrow ( \ppDS, \ppSC)$, return $\ppRS$.\\
~~~$\ppRS$ defines message space $\Mpp:= \Spp$.\\
\\
Algorithm $\RSKeyGen (\ppRS, 1^{\ell}):$\\
~~~$\ckSC \leftarrow \SCKeyGen(\ppSC, 1^{\ell})$, $(\pkDS, \skDS) \leftarrow \DSKeyGen(\ppDS)$, $\pkRS \leftarrow (\pkDS, \ckSC)$, $\skRS \leftarrow (\skDS, \ckSC)$.\\ 
~~~Return $(\pkRS, \skRS)$.\\
\\
Algorithm $\RSSign(\ppRS =(\skDS, \ckSC), \skRS, M):$\\
~~~$(C, O) \leftarrow \SCCommit(\ppSC, \ckSC, M)$, $\sigma_{C} \leftarrow \DSSign(\ppDS, \skDS, C)$.\\
%~~~Parse $\skRS$ as $(\skDS, \ckSC)$, $(C, O) \leftarrow \SCCommit(\ppSC, \ckSC, M)$, $\sigma_{C} \leftarrow \DSSign(\ppDS, \skDS, C)$.\\
~~~Return $\sigma \leftarrow (C, \sigma_{C}, O)$.\\
\\
Algorithm $\RSRedact(\ppRS = (\pkDS, \ckSC), \pkRS, M, \sigma = (C, \sigma_{C}, O), M'):$\\
%~~~Parse $\pkRS$ as $(\pkDS, \ckSC)$, $\sigma$ as $(C, \sigma_{C}, O)$.\\
~~~If $\DSVerify(\ppDS, \pkDS, C, \sigma_{C}) = 0$, return $\bot$.\\
~~~$W \leftarrow \SCOpenSubset(\ppSC, \ckSC, C, M, O, M')$, return $\sigma' \leftarrow (C, \sigma_{C}, W)$\\
\\

Algorithm $\RSVerify(\ppRS, \pkRS = (\pkDS, \ckSC), M, \sigma = (C, \sigma_{C}, O) ):$\\
%~~~Parse $\pkRS$ as $(\pkDS, \ckSC)$, $\sigma$ as $(C, \sigma_{C}, O)$.\\
~~~If $\DSVerify(\ppDS, \pkDS, C, \sigma_{C}) = 0$, return $0$.\\
~~~If $(\ppSC, C, M, O)$ is an input form of $\SCOpen$,\\
~~~~~~If $\SCOpen(\ppRS, C, M, O) = 1$, return $1$.\\
~~~If $(\ppSC, C, M, O)$ is an input form of $\SCVerifySubset$,\\
~~~~~~If $\SCVerifySubset(\ppSC, \ckSC, C, M, O) = 1$, return $1$.\\
~~~Otherwise return $0$.\\

\hline

\end{tabular}
\caption{\small
The construction of $\RSO$. 
}
\label{RSOframe}
\end{figure}

Clearly, the correctness of $\RSO$ is followed by that of $\DS$ and $\SC$.
The compactness of $\RSO$ is followed by that of $\SC$.

\subsection{Security Analysis} \label{SAnalysis}

\begin{theorem}\label{RSOunf}
If $\DS$ is EUF-CMA secure and $\SC$ is binding and subset-sound, then $\RSO$ is unforgeable.
\end{theorem}
Here, we give a sketch of the security proof. 
To explain the outline of the proof, we introduce new notations.
Let $q_s$ be the total number of queries from an adversary to $\mathcal{O}^{\Sign}$, $M_i$ be an $i$-th input for $\mathcal{O}^{\Sign}$, and $\sigma_i = (C_i, \sigma_{C_i}, O_i)$ be an $i$-th output of $\mathcal{O}^{\Sign}$.
We denote $Q^{C}_{\Sign} := \bigcup_{i=1}^{q_s} \{C_i\}$ and $Q^{M}_{\Sign} := \bigcup_{i=1}^{q_s} \{M_i\}$.
We consider three types of PPT adversaries $\A_1$, $\A_2$, and $\A_3$ that break the unforgeability security for $\RSO$.

\begin{itemize}
\item $\A_1$ generates a new commitment $C^* \notin Q^{C}_{\Sign}$, forges a signature $\sigma_{C^*}$ for $C^*$, and outputs a valid forgery $(m^* , \sigma^* = (C^*, \sigma_{C^*}, \allowbreak O^*))$.
That is, $\A_1$ does not reuse commitments output by $\mathcal{O}^{\Sign}$.
By the EUF-CMA security of $\DS$, it is difficult for $\A_1$ to forge a signature $\sigma_{C^*}$ for $C^* \notin Q^{C}_{\Sign}$.
Therefore, it is difficult for $\A_1$ to output a valid forgery $(M^* , \sigma^* = (C^*, \sigma_{C^*}, O^*))$. 
In the security proof, we construct $\B_1$ which breaking the EUF-CMA security of $\DS$ by using~$\A_1$.

\item $\A_2$ reuses a commitment $C^* \in Q^{C}_{\Sign}$ output by $\mathcal{O}^{\Sign}$, forges opening information $O^{*}$ for opening a message $M^*$ against $C^*$ where there is no $M \in Q^{M}_{\Sign}$ such that $M^* \subseteq M$, and outputs a forgery $(M^* , \sigma^* = (C^*, \sigma_{C^*}, \allowbreak O^*))$ where $C^* \in Q^{C}_{\Sign}$ and $\SCOpen(\ppSC, \ckSC, C^*, \allowbreak M^*, O^*) = 1$.
Since $\SC$ is bind, it is difficult for $\A_2$ to forge opening information $O^*$ for opening $M^*$ against $C^* \in Q^{C}_{\Sign}$.
In the security proof, we construct $\B_2$ which breaking the binding property of $\SC$ by using~$\A_2$.

\item $\A_3$ reuses a commitment $C^* \in Q^{C}_{\Sign}$ output by $\mathcal{O}^{\Sign}$, forges opening information $O^{*}$ for opening a message $M^*$ against $C^*$ where there is no $M \in Q^{M}_{\Sign}$ such that $M^* \subseteq M$, and outputs a forgery $(M^* , \sigma^* = (C^*, \sigma_{C^*}, \allowbreak O^*))$ where $C^* \in Q^{C}_{\Sign}$ and $\SCVerifySubset(\ppSC, \ckSC, \allowbreak C^*, M^*, O^*) = 1$.
Since $\SC$ is subset-sound it is difficult for $\A_3$ to forge opening information $O^*$ for opening $M^*$ against $C^* \in Q^{C}_{\Sign}$.
In the security proof, we construct $\B_3$ which breaking the subset-sound property of $\SC$ by using~$\A_3$.
\end{itemize}
Note that three types of forgers $\A_1$, $\A_2$, and $\A_3$ cover all the possibilities of forger's behaviors.
By constructing $\B_1$, $\B_2$, and $\B_3$, we prove the unforgeability security for $\RSO$.

\begin{theorem}\label{ROShidtheor}
If $\SC$ is perfectly hiding, then $\RSO$ is private.
\end{theorem}
{\bf Proof.} We consider the view of an adversary $\A$ in the privacy game.
To simplify the discussion, we assume that $\A$ queries $(M_{0}, M_{1}, M')$ to $\mathcal{O}^{\LR}$ where $M' \nsubseteq (M_{0} \cap M_{1})$.
Let $\sigma_b = (C_b, \sigma_{C}, O_b) \leftarrow \RSSign(\ppRS,\skRS, M)$ where $(C_b, O_b) \leftarrow \SCCommit(\ppSC, \ckSC, \allowbreak M_b)$, $\sigma_{C_b} \leftarrow \DSSign(\ppDS, \skDS, C_b)$ and $\sigma'_{b} = (C_b, \sigma_{C_b}, W_b) \allowbreak \leftarrow \RSRedact(\ppRS, \allowbreak \pkRS,\allowbreak M_b, \sigma, M')$, $W_b \leftarrow \SCOpenSubset(\ppSC, \allowbreak \ckSC, C_b, M_b, O_b, \allowbreak M')$ for $b \in \{0, 1\}$.

Now, we discuss distributions $\{\sigma'_{0} = (C_0, \sigma_{C_0}, W_0)\}$ and $\{ \sigma'_{1} = (C_1, \sigma_{C_1}, W_1)\}$ output by $\mathcal{O}^{\LR}$.
By the perfect hiding property (in Definition \ref{SChiddef}) of $\SC$, distributions $(C_0, W_0)$ and $(C_1, W_1)$ are identical.
Hence, the distributions $\{\sigma'_{0} = (C_0, \sigma_{C_0}, W_0)\}$ and $\{\sigma'_{1} = (C_1, \sigma_{C_1}, W_1)\}$ are identical.
From the above discussion, $\Adv^{\Priv}_{\RSO, \A} = 0$ holds. 
Therefore, we can conclude Theorem \ref{ROShidtheor}.
\qed

\section{Discussion}
We construct a redactable signature scheme with compact for sets.
If redactable signature scheme for sets is used as it is, there is a problem in real scenario.
For example, we consider the following submessages:
$m_1 = \mathrm{We}$, $m_2 =$ mustn't, $m_3 = \mathrm{go.}$, $m_4 = \mathrm{must}$, $m_5 = \mathrm{wait.}$.
If the adversary obtains a signature $\sigma$ on $M =\{$We mustn't go. We wait.$\}$, then the adversary re-orders it to $M' = \{$We must go. We mustn't wait.$\}$ and generates the proper signature on it.

Re-ordering message can be easily avoided by concatenating each submessage elements with an order-ID.
For instance, in above example, we change the submessages to $m_1 = \mathrm{We}||1$, $m_2 =$ mustn't$||2$, $m_3 = \mathrm{go.}||3$, $m_4 = \mathrm{We}||4$, $m_5 = \mathrm{must}||5$, $m_6 = \mathrm{wait.}||6$.
By concatenating each submessage with an order-ID, our redactable signature scheme is converted into the redactable signature scheme for lists and we can prevent re-ordering of submesssages.

\section*{Acknowledgement}
A part of this work was supported by a grant of Input Output Hong Kong, Nomura Research Institute, NTT Secure Platform Laboratories, Mitsubishi Electric, I-System, JST CREST JPMJCR14D6, JST OPERA and JSPS KAKENHI 16H01705, 17H01695.
We would also like to thank anonymous referees for their constructive comments.

\bibliographystyle{abbrvurl}
\bibliography{refRC.bib}

\appendix

\section{Security Proof for Theorem \ref{RSOunf}} \label{RSOunfappen}
An outline of the proof is described in Section \ref{SAnalysis}. 
Now, we prove Theorem \ref{RSOunf}.
 
{\bf Proof.} We consider the three types of adversaries $\A_1$, $\A_2$, and $\A_3$ described as follows and evaluate the advantage $\Adv^{\Uf}_{\RSO, \A_i}$ for each $i = 1, 2, 3$.

We consider an adversary $\A_1$ that generates a new commitment $C^* \notin Q^{C}_{\Sign}$, forges a signature $\sigma_{C^*}$ for $C^*$, and outputs a valid forgery $(M^* , \sigma^* = (C^*, \sigma_{C^*}, \allowbreak O^*))$.
We construct $\B_1$ which breaking the EUF-CMA security of $\DS$ by using $\A_1$ as follows.

\begin{itemize}
\item {\bf Initial setup:}
Given an input $(\ppDS, \pkDS)$ from the challenger of the EUF-CMA security game for $\DS$, $\B_1$ performs the following procedure.
\begin{itemize}
\item $\ppSC \leftarrow \SCSetup(1^\lambda)$, $\ckSC \leftarrow \SCKeyGen(\ppSC, 1^{\ell})$, $\ppRS \leftarrow(\ppDS, \ppSC)$,
$\pkRS \leftarrow (\pkDS, \ckSC)$, $Q^{C}_{\Sign} \leftarrow \{\}$, $Q^{M}_{\Sign} \leftarrow \{\}$.
\item Give $(\ppRS, \pkRS)$ to $\A_1$ as an input.\\
\end{itemize}

\item $\mathcal{O}^{\Sign}(M_i):$
Given an input $M_i$, $\B_1$ performs the following procedure.
\begin{itemize}
\item $(C_i, O_i) \leftarrow \SCCommit(\ppSC, \ckSC, M_i)$.
\item Query the challenger for the signature on the message $C_i$ and get its signature $\sigma_{C_i}$
\item $Q^{C}_{\Sign} \leftarrow Q^{C}_{\Sign} \cup \{C_i\}$, $Q^{M}_{\Sign} \leftarrow Q^{M}_{\Sign} \cup \{M_i\}$.
\item Return $(C_i, \sigma_{C_i}, O_i)$.\\
\end{itemize}

\item {\bf Output procedure:} 
$\B_1$ receives a forgery $(M^*, \sigma^*)$ output by $\A_1$.
Then $\B_1$ proceeds as follows.
\begin{enumerate}
\item \label{RSVreqO} $\RSVerify(\ppRS, \pkRS, M^*, \sigma^*) = 0$, then abort.
\item \label{RSMreqO} If there exists $M \in Q^{M}_{\Sign}$ such that $M^* \subseteq M$, then abort.
\item Parse $\sigma^*$ as $(C^*, \sigma_{C^*}, O^*)$.
\item \label{RSCnoRO}If $C^* \in Q^{C}_{\Sign}$, then abort.
\item Return $(C^*, \sigma_{C*})$ to the challenger.
\end{enumerate}
\end{itemize}

It is easy to see that $\B_1$ can simulate the unforgeability game for $\RSO$.
Now, we confirm that when $\A_1$ successfully output a valid forgery $(M^*, \sigma^*)$, $\B_1$ can forge a signature for $\DS$.
If $\A_1$ successfully output a valid forgery $(M^*, \sigma^*)$, $\B_1$ does not abort in Step \ref{RSVreqO}, \ref{RSMreqO} and \ref{RSCnoRO} of {\bf Output procedure}.
$\RSVerify(\ppRS, \pkRS, \allowbreak M^*, \sigma^*) = 1$ implies that $\DSVerify(\ppDS, \pkDS, C^*, \sigma_{C^*}) = 1$ holds.
Moreover, $\A_1$ outputs $C^* \notin Q^{C}_{\Sign}$.
This means that $\B_1$ does not make singing query $C^*$ to the challenger.
Therefore, $(C^*, \sigma_{C^*})$ is a valid forgery for $\DS$.

Finally, we evaluate the probability that $\B_1$ succeeds in forging a signature for $\DS$.
Let $\Adv^{\Uf}_{\RSO, \A_1}$ be the advantage of the unforgeability game for $\RSO$ of~$\A_1$.
The advantage of the EUF-CMA game for $\DS$ of $\B_1$ is 
\begin{equation}\label{UfRSOFup}
\Adv^{\sfEUFCMA}_{\DS, \B_1} \geq \Adv^{\Uf}_{\RSO, \A_1}.
\end{equation}

We consider an adversary $\A_2$ that reuses a commitment $C^* \in Q^{C}_{\Sign}$ output by $\mathcal{O}^{\Sign}$, forges opening information $O^{*}$ for opening a message $M^*$ against $C^*$ where there is no $M \in Q^{M}_{\Sign}$ such that $M^* \subseteq M$, and outputs a forgery $(M^* , \sigma^* = (C^*, \sigma_{C^*}, \allowbreak O^*))$ where $C^* \in Q^{C}_{\Sign}$ and $\SCOpen(\ppSC, \ckSC, C^*, M^*, O^*) = 1$.
We construct $\B_2$ which breaking the binding property of $\SC$ by using $\A_2$ as follows.

\begin{itemize}
\item {\bf Initial setup:}
Given an input $(\ppSC,\ckSC)$ from the challenger of the binding security game for $\SC$, $\B_2$ performs the following procedure.
\begin{itemize}
\item $\ppDS \leftarrow \DSSetup(1^\lambda)$, $(\pkDS, \skDS) \leftarrow \DSKeyGen(\ppDS)$, $\ppRS \leftarrow (\ppDS,  \allowbreak \ppSC)$, $\pkRS \leftarrow (\pkDS, \ckSC)$, $Q^{C}_{\Sign} \leftarrow \{\}$, $Q^{M}_{\Sign} \leftarrow \{\}$, $Q^{M, C, O}_{\Sign} \leftarrow \{\}$.
\item Give $(\ppRS, \pkRS)$ to $\A_2$ as an input.\\
\end{itemize}

\item $\mathcal{O}^{\Sign}(M_i):$
Given an input $M_i$, $\B_2$ performs the following procedure.
\begin{itemize}
\item $(C_i, O_i) \leftarrow \SCCommit(\ppSC, \ckSC, M_i)$.
\item $\sigma_{C_i} \leftarrow \DSSign(\ppDS, \skDS, \allowbreak C_i)$.
\item $Q^{C}_{\Sign} \leftarrow Q^{C}_{\Sign} \cup \{C_i\}$, $Q^{M}_{\Sign} \leftarrow Q^{M}_{\Sign} \cup \{M_i\}$, \\$Q^{M, C, O}_{\Sign} \leftarrow Q^{M, C, O}_{\Sign} \cup \{(M_i, C_i, O_i)\}$.
\item Return $(C_i, \sigma_{C_i}, O_i)$.\\
\end{itemize}

\item {\bf Output procedure:} 
$\B_2$ receives a forgery $(M^*, \sigma^*)$ output by $\A_2$.
Then $\B_2$ proceeds as follows.
\begin{enumerate}
\item \label{RSVreqTo} $\RSVerify(\ppRS, \pkRS, M^*, \sigma^*) = 0$, then abort.
\item \label{RSMreqTo} If there exists $M \in Q^{M}_{\Sign}$ such that $M^* \subseteq M$, then abort.
\item Parse $\sigma^*$ as $(C^*, \sigma_{C^*}, O^*)$.
\item \label{RSCnoRTo}If $C^* \notin Q^{C}_{\Sign}$, then abort.
\item Retrieve an entry $(M', C', O')$ from $Q^{M, C, O}_{\Sign}$ such that $C' = C^*$.  
\item Return $(C^*, M', O', M^*, O^*)$ to the challenger.
\end{enumerate}
\end{itemize}

It is easy to see that $\B_2$ can simulate the unforgeability game for $\RSO$.
Now, we confirm that when $\A_2$ successfully output a valid forgery $(M^*, \sigma^*)$, $\B_2$ can output a valid tuple $(C^*, M', O', M^*, O^*)$ for the binding game for $\SC$.
If $\A_2$ successfully output a valid forgery $(M^*, \sigma^*)$, $\B_2$ does not abort in Step \ref{RSVreqTo}, \ref{RSMreqTo} and \ref{RSCnoRTo} of {\bf Output procedure}.
By the strategy of $\A_2$, $\SCOpen(\ppSC, \ckSC, C^*, M^*, O^*) \allowbreak = 1$ holds.
$C^* \in Q^{C}_{\Sign}$ implies that there exists an entry $( M', C', O') \in Q^{M, C, O}_{\Sign}$ such that $C' = C^*$.
Moreover, since $\SC$ is correct, $\SCOpen(\ppSC, \ckSC, \allowbreak C'= C^*, M', O') = 1$ holds.
Furthermore, the fact that the $\B_2$ does not abort in Step \ref{RSMreqTo} of {\bf Output procedure} implies that $M^* \neq M'$.
Therefore, $(C^*, M', O', M^*, \allowbreak O^*)$ is a valid tuple for the binding game for $\SC$.

Finally, we evaluate the probability that $\B_2$ succeeds in outputting a valid tuple in the binding game for $\SC$.
Let $\Adv^{\Uf}_{\RSO, \A_2}$ be the advantage of the unforgeability game for $\RSO$ of~$\A_2$.
The advantage of the binding game for $\SC$ of $\B_2$~is 
\begin{equation}\label{UfRSOSup}
\Adv^{\Bind}_{\SC, \B_2} \geq \Adv^{\Uf}_{\RSO, \A_2}.
\end{equation}

We consider an adversary $\A_3$ that reuses a commitment $C^* \in Q^{C}_{\Sign}$ output by $\mathcal{O}^{\Sign}$, forges opening information $O^{*}$ for opening a message $M^*$ against $C^*$ where there is no $M \in Q^{M}_{\Sign}$ such that $M^* \subseteq M$, and outputs a forgery $(M^* , \sigma^* = (C^*, \sigma_{C^*}, \allowbreak O^*))$ where $C^* \in Q^{C}_{\Sign}$ and $\SCVerifySubset(\ppSC, \ckSC, \allowbreak C^*, M^*, O^*) = 1$.
We construct $\B_3$ which breaking the subset-sound property of $\SC$ by using $\A_3$ as follows.

\begin{itemize}
\item {\bf Initial setup:}
Given an input $(\ppSC,\ckSC)$ from the challenger of the subset-soundness security game for $\SC$, $\B_3$ performs the same procedure as {\bf Initial setup} of~$\B_2$.\\

\item $\mathcal{O}^{\Sign}(M_i):$
Given an input $M_i$, $\B_3$ performs the same procedure as $\mathcal{O}^{\Sign}(M_i)$ of $\B_2$.\\

\item {\bf Output procedure:} 
$\B_2$ receives a forgery $(M^*, \sigma^*)$ output by $\A_2$.
Then $\B_2$ proceeds as follows.
\begin{enumerate}
\item \label{RSVreqTr} $\RSVerify(\ppRS, \pkRS, M^*, \sigma^*) = 0$, then abort.
\item \label{RSMreqTr} If there exists $M \in Q^{M}_{\Sign}$ such that $M^* \subseteq M$, then abort.
\item Parse $\sigma^*$ as $(C^*, \sigma_{C^*}, O^*)$.
\item \label{RSCnoRTr}If $C^* \notin Q^{C}_{\Sign}$, then abort.
\item Retrieve an entry $(M', C', O')$ from $Q^{M, C, O}_{\Sign}$ such that $C' = C^*$.  
\item Return $(C^*, M', O', M^*, O^*)$ to the challenger.
\end{enumerate}
\end{itemize}

It is easy to see that $\B_3$ can simulate the unforgeability game for $\RSO$.
Now, we confirm that when $\A_3$ successfully output a valid forgery $(M^*, \sigma^*)$, $\B_3$ can output a valid tuple $(C^*, M', O', M^*, O^*)$ for the subset-soundness game for $\SC$.
If $\A_3$ successfully output a valid forgery $(M^*, \sigma^*)$, $\B_3$ does not abort in Step \ref{RSVreqTr}, \ref{RSMreqTr} and \ref{RSCnoRTr} of {\bf Output procedure}.
By the strategy of $\A_3$, $\SCVerifySubset(\ppSC, \ckSC, \allowbreak C^*, M^*, O^*) \allowbreak = 1$ holds.
When $C^* \in Q^{C}_{\Sign}$ holds then there exists an entry $(M', C', O') \in Q^{M, C, O}_{\Sign}$ such that $C' = C^*$.
Moreover, since $\SC$ is correct, $\SCOpen(\ppSC, \ckSC, \allowbreak C'= C^*, M', O') = 1$ holds.
Furthermore, the fact that the $\B_2$ does not abort in Step \ref{RSMreqTr} of {\bf Output procedure} implies that $M^* \nsubseteq M'$.
Therefore, $(C^*, M', O', M^*, \allowbreak O^*)$ is a valid tuple for the binding game for $\SC$.

Finally, we evaluate the probability that $\B_3$ succeeds in outputting a valid tuple in the subset-soundness game for $\SC$.
Let $\Adv^{\Uf}_{\RSO, \A_3}$ be the advantage of the unforgeability game for $\RSO$ of~$\A_3$.
The advantage of the subset-soundness game for $\SC$ of $\B_3$ is 
\begin{equation}\label{UfRSOTup}
\Adv^{\Sound}_{\SC, \B_3} \geq \Adv^{\Uf}_{\RSO, \A_3}.
\end{equation}

From inequalities (\ref{UfRSOFup}), (\ref{UfRSOSup}), and (\ref{UfRSOTup}), we can conclude Theorem \ref{RSOunf}. 
\qed% 

\section{Structure-Preserving Signature Scheme by Kiltz et al. \cite{KPW15} and Set-Commitment Scheme by Fuchsbauer et al. \cite{FHS19}}\label{SetComFHS}
In this section, we review bilinear groups.
Then, we review the structure-preserving signature scheme by Kiltz et al. \cite{KPW15} and the set-commitment scheme by Fuchsbauer et al. \cite{FHS19}.

\subsection{Bilinear Groups}
Let $\mathcal {G}$ be a bilinear group generator that takes as an input a security parameter $1^{\lambda}$ and outputs a descriptions of bilinear groups $\BG:=(\gk)$ where $\G_1$, $\G_2$ are additive groups of prime order $q$, $\G_T$ is a multiplicative group of prime order $q$, $e$ is an efficient computable, non-degenerating bilinear map $e:\G_1 \times \G_2 \rightarrow \G_T$, and $G_1$ and $G_2$ are generators of the group $\G_1$ and $\G_2$ respectively.
\begin{enumerate}
\item Bilinear: For all $a, b \in \Z_{q}$, then $e(aG_1, bG_2) = e(G_1, G_2)^{ab} = e(bG_1, aG_2)$.
\item Non-degenerate: $e(G_1, G_2) \neq 1_{\G_T}$. (i.e., $e(G_1, G_2)$ is a generator of $\G_T$.)
\end{enumerate}
We consider type 3 pairings where $\G_1 \neq \G_2$ and there are no efficiently computable homomorphisms between $\G_1$ and~$\G_2$.

We use the implicit representation of group elements by Escala, Herold, Kiltz, R{\`{a}}fols, and Villar \cite{EHKRV13}.
For $s \in \{1, 2\}$ and $a \in \mathbb{Z}_q$, we define $[a]_s := aG_s \in \G_s$ as the implicit representation of $a$ in $\G_s$.
For $s = T$ and $a \in \mathbb{Z}_q$, we define $[a]_T := e(G_1, G_2)^{a}\in \G_T$ as the implicit representation of $a$ in $\G_T$.
For a matrix $A = (a_{i,j}) \in \mathbb{Z}^{m \times n}_q$ we define $[A]_1$ as
\begin{equation*}
[A]_1 := \left(
 \begin{array}{ccc}
 a_{1,1}G_1 & \cdots & a_{1,n}G_1 \\
 \vdots & \ddots & \vdots \\
 a_{m,1}G_1 & \cdots & a_{m,n}G_1\\
 \end{array}
\right) \in \G_1^{m\times n}
\end{equation*}
and similarly for $[A]_2 \in \G_2^{m\times n}$ with a generator $G_2$, and $[A]_T \in \G_T^{m\times n}$ with a generator $e(G_1, G_2)$.

\begin{definition}[DDH Assumption in $\G_1$]\label{DDHasumdef}
Let $\mathcal{G}$ be a bilinear group generator.
The decisional Diffie-Hellman $(\DDH)$ assumption holds in $\G_1$ for $\mathcal {G}$ if for all PPT adversaries $\A$, the following advantage
\begin{equation*}
\begin{split} 
&\Adv^{\sfDDHO}_{\mathcal {G},\A}:= \Pr \left| \left[ b' = b \middle| 
\begin{split} 
&\BG=(q, \allowbreak \G_1, \allowbreak \G_2, \allowbreak \G_T, \allowbreak e,  G_1, \allowbreak G_2) \leftarrow \mathcal{G}(1^{\lambda}),\\
& x, y, z \xleftarrow{\$} \mathbb{Z}_q, b \xleftarrow{\$} \{0, 1\}, b' \leftarrow \A (\BG, [x]_1, [y]_1, [bxy +(1-b)z]_1 )
\end{split}
 \right] - \frac{1}{2}\right|
\end{split} 
\end{equation*}
is negligible in $\lambda$.
\end{definition}
Note that in the case of $b = 1$, $\A$ receives a Diffie-Hellman tuple $([x]_1, [y]_1, [xy]_1)$ as an input.
Similarly, in the case of $b =0 $, $\A$ receives a random tuple $([x]_1, [y]_1, \allowbreak [z]_1)$ as an input.

The dual of the above assumption is the Decisional Diffie-Hellman assumption in $\G_2$ for $\mathcal {G}$, which is defined by changing the roles of $\G_1$ to $\G_2$ in Definition~\ref{DDHasumdef}. 

\begin{definition}[SXDH Assumption \cite{BGMM05}]
Let $\mathcal{G}$ be a bilinear group generator outputting $\BG:=(\gk)$.
The symmetric external Diffie-Hellman $(\SXDH)$ assumption holds for $\mathcal{G}$ if the $\DDH$ assumption holds both $\G_1$ and $\G_2$.
\end{definition}

\begin{definition}[$q$-co-DL Assumption \cite{FHS19}]
Let $\mathcal{G}$ be a bilinear group generator.
The $q$-co-discrete logarithm $(\qcoDL)$ assumption holds for $\mathcal {G}$ if for all PPT adversaries $\A$, the following advantage
\begin{equation*}
\begin{split}
&\Adv^{\sfqcoDL}_{\mathcal {G},\A}:= \Pr \left[ a' = a \middle| 
\begin{split} 
&\BG \leftarrow \mathcal{G}(1^{\lambda}), a \xleftarrow{\$} \mathbb{Z}_q, a' \leftarrow \A \left(\BG, ([a^{j}]_1, [a^{j}]_2)_{j \in [q]} \right)
\end{split}
 \right]
 \end{split}
\end{equation*}
is negligible in $\lambda$.
\end{definition}

\begin{definition}[$q$-co-GSDH Assumption \cite{FHS19}]
Let $\mathcal{G}$ be a bilinear group generator.
The $q$-co-generalized-strong-Diffie-Hellman $(\qcoGSDH)$ assumption holds over $\mathcal {G}$ if for all PPT adversaries $\A$, the following advantage
\begin{equation*}
\begin{split} 
&\Adv^{\sfqcoGSDH}_{\mathcal {G},\A}:= \Pr \left[ 
\begin{split} 
& T \in \G_1 \land g, h \in \mathbb{Z}_q[X]\\
& \land 0 \leq \deg g < \deg h \leq q \\
& \land e(T, [h(a)]_2) = e([g(a)]_1, [1]_2)
\end{split} 
\middle| 
\begin{split} 
&\BG \leftarrow \mathcal{G}(1^{\lambda}), a \xleftarrow{\$} \mathbb{Z}_q,\\
&\begin{split} &(g, h, T) \leftarrow \A (\BG, ([a^{j}]_1, [a^{j}]_2)_{j \in [q]} ) \end{split} 
\end{split}
 \right]
\end{split} 
\end{equation*}
is negligible in $\lambda$.

\end{definition}

\subsection{Structure-Preserving Signature Scheme by Kiltz et al.}

We review the structure-preserving signature by Kiltz et al.~\cite{KPW15}.
This scheme is efficient and its security is proven without GGM and supports a multi-message (vector message) signing.
In this work, we only need a single-message signing scheme. 
Now, we describe the structure-preserving signature scheme $\DS_{\KPW}$ given by Kiltz et al. in Fig.~\ref{KPWSPS}.

\begin{figure}[htbp]
\centering
\begin{tabular}{|l|}
\hline
Algorithm $\SPSSetup_{\KPW}(1^{\lambda}):$\\
~~~$\BG:=(\gk) \leftarrow \mathcal{G}(1^{\lambda})$, return $\ppDS \leftarrow \BG$.\\
~~~$\ppDS$ defines message space $\MppDS ;= \mathbb{G}_1$.\\
\\
Algorithm $\SPSKeyGen_{\KPW}(\ppDS):$\\
~~~$a, b \xleftarrow{\$} \mathbb{Z}_{q}$, $K \xleftarrow{\$} \mathbb{Z}_q^{2 \times 2}$, $A \leftarrow (1, a)^{\top} \in \mathbb{Z}_q^{2 \times 1}$, $B \leftarrow (1, b)^{\top} \in \mathbb{Z}_q^{2 \times 1}$,\\
~~~$K_0, K_1 \xleftarrow{\$} \mathbb{Z}_q^{2 \times 2}$, $D \leftarrow KA$, $D_0 \leftarrow K_0A$, $D_1 \leftarrow K_1A$, $P_0 \leftarrow B^{\top}K_0$, $P_1 \leftarrow B^{\top}K_1$,\\
~~~$\pkDS \leftarrow ([D_0]_2, [D_1]_2, [D]_2, [A]_2)$, $\skDS \leftarrow ([D_0]_2, [D_1]_2, [D]_2, [A]_2, K, [P_0]_1, [P_1]_1, [B]_1)$.\\
~~~Return $(\pkDS, \skDS)$.\\
\\
Algorithm $\SPSSign_{\KPW}(\ppDS, \skDS, [m]_1):$\\
~~~Parse $\skDS$ as $([D_0]_2, [D_1]_2, [D]_2, [A]_2, K, [P_0]_1, [P_1]_1, [B]_1)$.\\
~~~$r, \tau \xleftarrow{\$} \mathbb{Z}_{q}$, $\theta_1 \leftarrow [(1, m)K + r (P_0 + \tau P_1)]_1 \in \G_1^{1 \times 2}$,\\
~~~$\theta_2 \leftarrow [rB^{\top}]_1 \in \G_1^{1 \times 2} $, $\theta_3 \leftarrow [rB^{\top} \tau]_1 \in \G_1^{1 \times 2}$, $\theta_4 \leftarrow [\tau]_2 \in \G_2$. \\
~~~Return $\sigma \leftarrow (\theta_1, \theta_2, \theta_3, \theta_4)$.\\
\\
Algorithm $\SPSVerify_{\KPW}(\ppDS, \pkDS, [m]_1, \sigma):$\\
~~~Parse $\pkDS$ as $([D_0]_2, [D_1]_2, [D]_2, [A]_2)$, $\sigma$ as $(\theta_1, \theta_2, \theta_3, \theta_4)$.\\
~~~If $e(\theta_1, [A]_2) = e([(1, m)]_1, [D]_2) e(\theta_2, [D_0]_2) e(\theta_3, [D_1]_2) \land e(\theta_2, \theta_4) = e(\theta_3. [1]_2)$, return $1$.\\
~~~Otherwise, return $0$.\\

\hline
\end{tabular}
\caption{\small The construction of $\SPS_{\KPW}$.}
\label{KPWSPS}

\end{figure}

 \begin{lemma}[\cite{KPW15}]\label{SPSKPWUf}
 If the $\SXDH$ assumption holds for $\mathcal {G}$, then $\SPS_{\KPW}$ is EUF-CMA secure \footnote{Kiltz et al. \cite{KPW15} proved that $\SPS_{\KPW}$ satisfies EUF-CMA security under the $\mathcal{D}_k$-matrix Diffie-Hellman ($\mathcal{D}_k$-MDDH) assumption \cite{EHKRV13}.
 If $k=1$, the $\mathcal{D}_1$-MDDH assumption corresponds to the SXDH assumption.
 In Lemma \ref{SPSKPWUf}, we rewrite the claim of Kiltz~et~al. in \cite{KPW15} as $k=1$.}.
 \end{lemma}

\subsection{Set-Commitment Construction by Fuchsbauer et al.}

Let $\mathcal {G}$ be a bilinear group generator which outputs a descriptions of multiplicative groups in type 3 pairings and $\ell$ be the upper bound for the number of elements of sets to be signed.
For a non-empty set $S = \{s_1, \dots s_{\#S}\} \subseteq Z_{q}$, we define the polynomials $f_{S}(X) := \prod_{s \in S} (X- s) = \sum^{\#S}_{i=0}f_{i} \cdot X^{i}$ where $f_0 = (-1)^{\#S}(s_1s_2\dots s_{\#S})$, $\dots$, $\allowbreak f_{\#S-1} = (-1)(s_1 + \dots + s_{\#S})$ and $f_{\#S} = 1$.
$[f_{S}(a)]_1$ can be computed from $S$ and $([a^{i}]_1)^{\#S}_{i=0}$ without using $a$.
In the case of $S = \emptyset$, we define $f_{\emptyset}:= 1$.

Now, we describe the set-commitment scheme $\SC_{\FHS}$ given by Fuchsbauer et al. \cite{FHS19} in Fig.~\ref{FHSSC}.
\begin{figure}[htbp]
\centering
\begin{tabular}{|l|}
\hline
Algorithm $\SCSetup_{\FHS}(1^{\lambda}):$\\
~~~$\BG:=(\gk) \leftarrow \mathcal{G}(1^{\lambda})$, return $\ppSC \leftarrow \BG$.\\
~~~$\ppSC$ defines message space $\Spp := \mathbb{Z}_q$.\\
\\
Algorithm $\SCKeyGen_{\FHS}(\ppSC, 1^{\ell}):$\\
~~~$a \xleftarrow{\$} \mathbb{Z}_{q}$, return $\ckSC \leftarrow (([a^{i}]_1, [a^{i}]_2)_{i \in [\ell]})$.\\
\\
Algorithm $\SCCommit_{\FHS}(\ppSC, \ckSC = (([a^{i}]_1, [a^{i}]_2)_{i \in [\ell]}), S \subseteq \Spp):$\\
%~~~Parse $\ckSC$ as $(([a^{i}]_1, [a^{i}]_2)_{i \in [\ell]})$.\\
~~~If $S \nsubseteq \mathbb{Z}_q \lor \#S = 0 \lor \ell < \#S $, return $\bot$.\\
~~~If there exist $a' \in S$ such that $[a']_1 = [a]_1$, $C \xleftarrow{\$} \mathbb{G}^*_1$, $O \leftarrow (1, a')$, return $(C, O)$. \\
~~~$\rho \xleftarrow{\$} \mathbb{Z}^*_q$, $C \leftarrow [\rho \cdot f_{S}(a)]_1$, $O \leftarrow (0, \rho)$, return $(C, O)$. \\
\\
Algorithm $\SCOpen_{\FHS}(\ppSC, \ckSC=(([a^{i}]_1, [a^{i}]_2)_{i \in [\ell]}), C, S, O=(b, \rho)):$\\
%~~~Parse $\ckSC$ as $(([a^{i}]_1, [a^{i}]_2)_{i \in [\ell]})$, $O$ as $(b, \rho)$.\\
~~~If $C \notin \mathbb{G}^*_1 \lor \rho \notin \mathbb{Z}_q$, return $0$.\\
~~~If $S \nsubseteq \mathbb{Z}_q \lor \#S = 0 \lor \ell < \#S $, return $\bot$.\\
~~~If $b= 1 \land [\rho]_1 = [a]_1$, return $1$. \\
~~~If $b= 0 \land C = [\rho \cdot f_{S}(a)]_1$, return $1$. \\
~~~Otherwise, return $0$.\\
\\
Algorithm $\SCOpenSubset_{\FHS}(\ppSC, \ckSC = (([a^{i}]_1, [a^{i}]_2)_{i \in [\ell]}), C, S, O = (b, \rho), S'):$\\
~~~$\SCOpen_{\FHS}(\ppSC, \ckSC, C, S, O) = 0 \lor S' \nsubseteq S \lor S = \emptyset$, return $\bot$.\\
%~~~Parse $\ckSC$ as $(([a^{i}]_1, [a^{i}]_2)_{i \in [\ell]})$, $O$ as $(b, \rho)$.\\
~~~If $b=1$,\\
~~~~~~If $\rho \in S'$, return $W \leftarrow \bot$, otherwise return $W \leftarrow f_{S'}(\rho)^{-1} \cdot C$.\\
~~~If $b=0$, return $W \leftarrow [\rho \cdot f_{S \backslash S'}(a)]_1$.\\
\\
Algorithm $\SCVerifySubset_{\FHS}(\ppSC, \ckSC=(([a^{i}]_1, [a^{i}]_2)_{i \in [\ell]}), C, T, W):$\\
%~~~Parse $\ckSC$ as $(([a^{i}]_1, [a^{i}]_2)_{i \in [\ell]})$.\\
~~~If $C \notin \mathbb{G}^*_1$, return $0$.\\
~~~If $T \nsubseteq \mathbb{Z}_q \lor \mathbb{Z}_q \lor \#T = 0 \lor \ell < \#T$, return $0$.\\
~~~If there exist $\rho'$ such that $[\rho']_1 = [a]_1$,\\
~~~~~~~If $W = \bot$, return $1$, otherwise, return $0$.\\
~~~If $W \in \mathbb{G}^*_1 \land e(W, [f_{T}(a)]_2) = e (C, [1]_2)$, return $1$.\\
~~~Otherwise, return $0$. 
\\
\hline

\end{tabular}
\caption{\small The construction of $\SC_{\FHS}$.}
\label{FHSSC}
\end{figure}
Regarding security for $\SC_{\FHS}$, the following facts were clarified in \cite{FHS19}.

\begin{lemma}[\cite{FHS19}]\label{FHSSCpc}
$\SC_{\FHS}$ is perfectly correct.
\end{lemma}

\begin{lemma}
$\SC_{\FHS}$ satisfies compactness.
\end{lemma}

\begin{lemma}[\cite{FHS19}]\label{FHSSCbin}
Let $\ell$ be an upper bound for the number of elements in committed sets.
If the $\ellcoDL$ assumption holds for $\mathcal {G}$, then $\SC_{\FHS}$ is binding.
\end{lemma}

\begin{lemma}[\cite{FHS19}]\label{FHSSCSubsou}
Let $\ell$ be an upper bound for the number of elements in committed sets.
If the $\ellcoGSDH$ assumption holds for $\mathcal {G}$, then $\SC_{\FHS}$ is subset-sound.
\end{lemma}

\begin{lemma}[\cite{FHS19}]\label{FHSSCSHid}
$\SC_{\FHS}$ is perfectly hiding.
\end{lemma}

\setcounter{tocdepth}{2}
\tableofcontents

\end{document}